**Potential multiple steady-states in the long-term carbon cycle**

Stephen Tennenbaum[1], Faina Berezovskaya [2], David Schwartzman[3*]

1 Department of Mathematics, George Washington University, Washington D.C., U.S.
2 Department of Mathematics, Howard University, Washington D.C. 20059, U.S.
3 Department of Biology, Howard University, Washington D.C. 20059, U.S.
* email: dschwartzman@gmail.com

**In our modeling of the long-term carbon cycle we find potential multiple steady-states in Phanerozoic climates. We include the effects of biotic enhancement of weathering on land, organic carbon burial, oxidation of reduced organic carbon in terrestrial sediments and the variation of biotic productivity with temperature, finding a second stable steady-state appearing between 20 and 50 ºC. The very warm early Triassic climate as well as an oceanic anoxic event in the late Cretaceous may be the potential candidates for an upper temperature steady-state. Given our results, the anthropogenic driven rise of atmospheric carbon dioxide could potentially push the climate into tipping points to a modestly higher temperature steady-state, instead of relaxing back to pre-anthropogenic conditions.**

For a given set of fluxes into and out of the Earth's surface system a single steady-state attractor in the long-term carbon cycle has been previously assumed in modeling the past climates in the Phanerozoic (Berner, 2004) and even deep in Precambrian time (Schwartzman, 1999 2002; Schwartzman, 2008). In their modeling of the lifespan of the biosphere Lenton and von Bloh (2001) found solutions for more than one stable steady-state for the future, corresponding to abiotic and biotic surfaces. Multiple steady-states have also been proposed for tectonic regimes in terrestrial planets (Lenardic and Crowley, 2012).

Here we explore potential multiple stable steady-states in Phanerozoic climates. This paper presents the results of systematic modeling suggested from previous studies (Schwartzman and Kleidon, 2006; Schwartzman and Berezovskaya, 2009).

Only the reaction of CaMg silicates with water and atmospherically-derived carbon dioxide results in a net transfer of carbon from the atmosphere to the crust, via the formation of bicarbonate and calcium/magnesium cations, their transfer to the ocean, and the precipitation of calcium/magnesium carbonate; the weathering of limestone produces no net change in carbon dioxide in the atmosphere/ocean system. In equation (1) $CaSiO_3$ is the proxy for more complex silicates, which include Mg as well as Ca, for simplicity Mg is omitted.

On land:
(1) $CaSiO_3 + H_2O + 2CO_2 \rightarrow 2HCO_3^- + Ca^{+2} + SiO_2$
In ocean:
(2) $2HCO_3^- + Ca^{+2} \rightarrow CaCO_3 + CO_2 + H_2O$

The reverse reaction from right to left in equation (2) is the weathering of limestone on land, hence only the weathering of Ca silicates corresponds to a net transfer of carbon dioxide from the atmosphere to the crust, noting that for each mole of $CaSiO_3$ reacting, there is a net consumption of one mole of $CO_2$, which is buried on the ocean floor as limestone.

Figure 1 illustrates the relevant fluxes in the long-term carbon cycle. All fluxes correspond to carbon dioxide transfer. V is the volcanic source flux, $F_3$ is the burial flux of reduced organic carbon, $F_4$ the oxidation flux of reduced organic carbon in sediments exposed on land. $F_1$ is the deposition flux of CaMg carbonates in the ocean, and $F_2$ corresponds to the weathering flux of carbonates on land. Since $F_1$ corresponds to the total flux of carbonate deposited in the ocean, it is derived from the reaction of atmospheric carbon dioxide with CaMg silicates and carbonates in weathering reactions on land. Thus ($F_1 - F_2$) corresponds to the CaMg silicate weathering flux alone ("$W_{sil}$") since a flux equal to $F_2$ is deposited as carbonate in the ocean, i.e., $W_{sil} = F_1 - F_2$.

The role of weathering of continental shelf marine sediments may be significant in the sequestration of atmospherically-derived carbon dioxide, contributing to $W_{sil}$ (Wallmann et al., 2008; Jeandel et al., 2011). Comprehensive models of the long-term carbon cycle should capture variation in mountain uplift and seafloor weathering processes (e.g., Berner and Kothavala, 2001), as well as paleogeography (Godderis et al. 2014).

The emergence of land plants in the early Phanerozoic introduced new processes into the surface system. Given the increasing dominance of higher plants in Phanerozoic time, a comprehensive model of the long-term carbon cycle should take into account the influences of the global biotic productivity as a function of $pCO_2$, temperature, and planetary albedo, along with the rate of oxidation of reduced organic carbon, sulfides, and ferrous iron in sediments, as well as the burial of reduced organic carbon in sediments, and interactions with the biogeochemical cycles of phosphorus (Berner, 1999; Beerling and Berner, 2005).

Biologic evolution has had a major influence on biotic productivity, the biotic enhancement of silicate weathering (BEW) and all the biogeochemical cycles (Schwartzman, 2008; Berner, 1999; Beerling and Berner, 2005; Taylor et al., 2012). The BEW is defined as how much faster the silicate weathering carbon sequestration flux is under biotic conditions than under abiotic conditions at the same surface temperature and $pCO_2$ level. Whereas this silicate weathering carbon flux with the present terrestrial biota is likely at least an order of magnitude greater than under abiotic conditions, the same flux can be achieved with abundant life at lower temperatures and $pCO_2$ levels than in its absence. Since the rise of higher plants,

BEW has included multifold processes, namely soil stabilization, elevated carbon dioxide levels in soil from root/microbial respiration and decay, the interactions in the rhizosphere/mycorrhizae, and other processes driven by biology (Berner, 2004; Schwartzman, 2008, 1999 2002).

**Modeling the Phanerozoic long-term carbon cycle**

As a first approximation, we simplify our model by ignoring the rate of oxidation of sulfides and ferrous iron in sediments as well as the potential variation in the planetary albedo, arising from land versus oceanic area and cloud variability, but include the kinetics of silicate weathering (with the influence of its biotic enhancement), organic carbon burial, oxidation of reduced organic carbon in terrestrial sediments, as well as the influence of the variation of biotic productivity with temperature on relevant processes. This modeling is informed by the current knowledge of the kinetics of each process. The greenhouse functions used that connect temperature to the atmospheric carbon dioxide level are shown in Figure 2. Modeling details, including greenhouse functions used, are found in Supplementary Information

The rate of change of atmospheric carbon dioxide, $pCO_2$, is equal to the difference between source and sink fluxes with respect to the atmosphere/ocean pool.

$$(3) \quad dpCO_2/dt = V - W_{sil} - F_3 + F_4$$

We define R as follows: $R = W_{sil} + F_3 - F_4$. All the terms making up R are functions of temperature; hence, R will vary as temperature increases from the assumed stable steady-state at 15 ºC to the maximum tolerated by plants, 50 ºC.

At steady-state $dpCO_2/dt = 0$, thus at steady-state the following equality is required

$$(4) \quad V = W_{sil} + F_3 - F_4 = R$$

where these $W_{sil}$, $F_3$, $F_4$, values are generated at the temperature/ $pCO_2$ levels at steady-state. Since in the Phanerozoic steady-state in the long-term carbon cycle is achieved in a $10^5$ to $10^6$ year time scale (1, 3), components that equilibrate on a shorter time scale, namely biota, soil, atmosphere and ocean, are subsumed in one reservoir, the atmosphere/ocean pool (Figure 1).

**Stable and unstable steady-states in the long-term carbon cycle**

Recall that $dpCO_2/dt = V - R$. Since the climate sensitivity is a monotonically increasing function (Figure 2), the rate of change of temperature is positively correlated to departure from steady-state, i.e. $dT/dt$ varies in the same direction with $V - R$, since the temperature is driven by the $pCO_2$ level as a consequence of greenhouse forcing (temperature corresponds to the global mean surface value). The latter is the critical mechanism generating stable steady-states in the long-term carbon cycle because of the role of temperature and $pCO_2$ levels in processes entailed in R.

In Figure 3a we plot V – R on the y-axis versus Temperature on the x-axis. When V > R the temperature (T) rises, since $dpCO_2/dt$ is positive, and when V < R the temperature (T) decreases since $dpCO_2/dt$ is negative. Steady-state occurs when V = R (equation (4)). A negative slope as the curve crosses the x-axis (when V = R) implies a stable steady-state, indicated by an "S" on the graph. A steady-state is locally stable if small deviations from a steady- state result in the system returning to it. If a slight decrease in temperature causes V > R and T increases in response, and if a slight increase in temperature causes V < R and T decreases then the steady-state is stable.

A steady-state is unstable if the opposite happens; a slight shift in temperature away from the steady-state is positively reinforced. This point is indicated by the "U" on the graph. These unstable steady-states mark a tipping point. Tipping points are increasingly recognized in the carbon cycle (Lenton, 2011; Lenton et al., 2008). They divide the system into two basins of attraction. Small perturbations of temperature and $pCO_2$ level away from the lower stable steady-state cause the system to remain in that basin and the system tends to return to the lower stable steady-state. Perturbations large enough to shift temperature and $pCO_2$ level past the tipping point will push the system into the basin for the upper stable steady-state and the system will shift to that steady-state. Note that the perturbation only needs to move the system past the unstable steady-state to move into the new basin, after that the system will move to the new stable steady-state on its own.

**Computing the stable and unstable steady-states**

R values were computed as a function of temperature up to the habitability limit of plants (50 ºC), with V held constant (e.g., Figure 3b). For most sensitivity tests we assume the following values at the assumed lower temperature stable steady-state corresponding to 15 ºC: V = 0.5, $W_{silo}$ = 0.5, $F_{3o} = F_{4o}$ = 0.1, hence $W_{silo}$ = V. Note that the values of $W_{sil}$, $F_3$, and $F_4$ at 15 ºC are labeled with subscript "o".

Modeling results are shown in Figures 3b, c, d, f, g with parameters B, α and b defined along with other modeling details in Supplementary Information.

The temperature corresponding to the second stable steady-state was also analyzed with respect to variation of the relative fluxes comprising R at the same lower temperature stable steady-state corresponding to 15 ºC, i.e., $W_{silo}$, $F_{3o}$, $F_{4o}$, again keeping V at the same value and held constant. Maximizing $F_{4o}$ with respect to $W_{silo}$ with $F_{3o}$ = 0 results in the lowest upper temperature steady-state ("UTSS"). Results are shown in Figure 3e, with modeling details given in Supplementary Information.

A discussion of the potential implications of these results to actual paleoclimates immediately follows below.

**Modeling results and potential paleoclimates with an upper temperature steady-state**

We note that the minimum UTSS is 38 °C for $W_{silo} = V$, with $W_{silo} = 0.5$, $F_{3o} = F_{4o} = 0.1$ (Figures 3b, c, d). A temperature of 38 °C corresponds to a very warm climate, but not too hot to support plant life, especially since the computed temperature is the global mean surface temperature. The corresponding values of BEW computed at 15 °C are 41 and 20 using greenhouse functions (9) and (10) respectively (Figures 3c, d). The minimum values of BEW computed at 15 °C are 23 and 14 using greenhouse functions (9) and (10) respectively, with a corresponding UTSS value of 41 °C. These BEW values are consistent with estimates for near modern global BEW (Schwartzman, 2008, 1999 2002).

Thus, these results of our modeling suggest that a very warm climate at steady-state in the long-term carbon cycle may have occurred since plants first dominated the land biota, some 400 million years ago. How could such a steady-state emerge? One possibility is that a transient volcanic episode injects sufficient carbon dioxide into the atmosphere to reach $pCO_2$ and temperature levels past the unstable steady-state. Once volcanic outgassing plunges down to the previous occurring levels the steady-state persists, until such time another perturbation, such as orbital-induced cooling in the Milankovitch cycle, pushes the climate system back into the lower temperature steady-state.

Are there plausible paleoclimates that correspond to the upper temperature steady-state implied by our modeling? Two possibilities are suggested, the mid-Cretaceous and high $pCO_2$ middle Eocene climates. Both climates may have been generated by volcanic episodes outgassing carbon dioxide to the atmosphere (Pagani et al., 2006; Bijl et al., 2010; Hu et al., 2012; Hong and Lee, 2012; Lee et al., 2013), although trace greenhouse gases such as methane may have contributed to these warm climates (Beerling et al., 2011). The mid-Cretaceous climate reached temperatures of roughly 35 °C at the equator, 25 °C at 60 N latitude (Hu et al., 2012), somewhat cooler than our model minimum upper temperature stable steady-state, which was computed without including the small influence of a slightly lower solar luminosity on the greenhouse function a hundred million years ago. However, the apparently very warm Early Triassic climate, with surface ocean temperatures of 36-40 °C (Sun et al., 2012) may be the prime candidate for this very warm steady-state.

We compute even lower UTSS temperatures by allowing the values of $F_{3o}$, $F_{4o}$ and $W_{silo}$ to vary, assuming a constant V, again producing a stable steady-state at 15 °C. When $F_{3o} = 0$ and $F_{4o} = 1$ then the lowest computed UTSS is reached at 20 °C (Figure 3e) with a corresponding value of BEW of 14 computed at 15 °C (using equations (9) and (13), Supplementary Information). Using equation (10) the minimum UTSS derived is 21 °C with BEW computed at 15 °C equal to 7 assuming the same limiting constants as in Figure 3e. Again these computed BEW values are consistent with estimates of near modern BEW.

A significant decrease in organic carbon burial may occur in a warmer ocean driving a higher rate of bacterial respiration as proposed for the Eocene (John et al., 2013), thereby creating the conditions for a lower temperature UTSS. A potential candidate for this UTSS is suggested from the modeling of the Oceanic Anoxic Event 2 in the late Cretaceous (Pogge von Strandmann et al., 2013). At the close of this Event the estimated amount of volcanic outgassed $CO_2$ apparently exceeds the amount sequestered by enhanced weathering (Pogge von Strandmann et al., 2013), leaving the climate in a higher temperature state than before the upsurge in volcanism.

Is the bimodal distribution of inferred $pCO_2$/temperature levels over Phanerozoic time evidence for multiple steady-states as proposed in this paper ? The early Paleozoic decrease in $pCO_2$ levels and the dramatic cooling of the Permo-Carboniferous glacial epoch (see e.g., Berner, 2004) might be considered as potential evidence, but the former is more plausibly explained by the rise of higher plants and the first forests (Berner and Kothavala, 2001), coupled with changes in weathering intensity induced by continental drift (Godderis et al., 2014), and the latter by enhanced burial of reduced organic carbon (coal), rather than being examples of the shifts driven by transient changes in the volcanic outgassing flux increasing or decreasing atmospheric $pCO_2$ levels and temperature levels along the lines of our modeling.

Geochemically derived estimates of the ratios of carbon sources and sinks along with paleotemperatures/ $pCO_2$ levels could help identify multiple steady-states existing in these ancient climates. Our model results suggest further exploration of potential upper temperature steady-state climates, including more variables, such as planetary albedo, variation in weatherability, as well as ice-cap albedo feedbacks and latitudinal temperature variation instead of using the global mean surface temperature.

Our modeling of the Phanerozoic assumes constant planetary albedo, similar to the present. But is this also true of potentially more extreme climate regimes such as in the early Archean or Hadean? We explore this possibility in Supplementary Information.

Another interesting possibility has emerged from our modeling. Previous studies of the carbon cycle have explored the long-term relaxation of the climate, given assumed inputs of fossil fuel carbon into the atmosphere (Archer et al., 2009). However, the anthropogenic derived rise of atmospheric carbon dioxide could drive the climate into tipping points corresponding to a modestly higher temperature steady-state if the biosphere is tuned to the appropriate parameters and kinetic behavior as explored in our study. Just as increased outgassing of volcanoes could produce a transient condition necessary to push the system past the critical (tipping) point, likewise continued unabated fossil fuel consumption could also push the system into the upper basin of the higher steady-state.

We conclude that multiple steady-state attractors in the long-term carbon cycle may have influenced climates of the past and played a role in stabilizing transient excursions. Further modeling along with paleoclimatic and geochemical studies should further illuminate the Earth's climatic history and its coupling to biotic evolution, including the possibility of multiple steady-states in the long-term carbon cycle.

**Acknowledgments**

We thank Tyler Volk and Axel Kleidon for their long discussions on the subject of this paper and previous collaboration.

**Author contributions**

S. T. and F.B. computed the model results. D.S. generated the conceptual framework for modeling. D.S. and S.T. wrote the manuscript with assistance of F.B.

**Supplementary Information**

**Modeling details**

**Parameterization of kinetics generating relevant fluxes**

Volcanic outgassing (V) is held constant in our modeling. Further studies should take into account potential variation in V as a result of changes in the partitioning of carbonate burial between the continental shelf and deep ocean floor and the resultant degassing of carbon dioxide in subduction (Volk, 1989; Caldeira, 1991), as well as changes in juvenile mantle degassing. We assume steady-state occurs at 15 ºC with R and its components $W_{sil}$, $F_3$, and $F_4$ labeled with subscript "o" at this temperature. For modeling results shown in Figures 3b, c, d, f, g we assume the following values: $V = 0.5$, $W_{silo} = 0.5$, $F_{3o} = F_{4o} = 0.1$. These assumed values are simple multiples of whatever unit of flux is assigned appropriate for the long-term carbon cycle, with their ratios approximating plausible geochemical estimates (Berner, 2004; Berner and Kothavala, 2001). The results of a sensitivity test for variation in the $F_{3o} / F_{4o}$ ratio is shown in Figure 3e. The ratio ($W_{sil}/ (F_3 – F_4)$) in the Phanerozoic has been roughly equal to five (Berner, 1991). In our modeling we test the sensitivity with respect to this ratio.

The silicate weathering flux in the presence of life is given as follows.

$$(5) \quad W_{sil} = A\,[BG^{0.4} + e^{0.073(T-15)}\,(Pr^{\alpha})],$$

where $A = 0.5 /(BG_o^{0.4} + 1)$, composed so that $W_{sil, o} = 0.5$ at the reference temperature of 15 °C. $Pr = (pCO_2 / pCO_{2\ reference})$ where the reference level corresponds to the pre-anthropogenic modern atmospheric carbon dioxide level, i.e., 0.0003 bar. . T is the global mean surface temperature in °C. G is defined as the global biological productivity, to be explained further for equation (6).

The $\alpha$ constant determines the influence of $pCO_2$ level on the silicate weathering flux, with assumed values ranging from 0.3 to 0.4 (Schwartzman,1999 2002). B is a variable multiplier of the G term; B is an output in the analysis of steady-states emerging between 15 and 50 °C. The $BG^{0.4}$ term corresponds to the biological driver of weathering, a function of BEW, with $G^{0.4}$ dependency (Volk, 1987; Caldeira and Kasting, 1992). The second term, $e^{0.073(T-15)}\,(Pr^{\alpha})$, is the abiotic weathering factor. The temperature influence on weathering is captured by the $e^{0.073(T-15)}$ factor, noting that the exponential constant 0.073 corresponds to the sum of the direct temperature effect on weathering factor, 0.056 plus the runoff factor 0.017 (Schwartzman,1999 2002).

The biotic enhancement of weathering (BEW) at 15 °C is defined as the ratio of biotic $W_{sil}$ / abiotic $W_{sil}$ (Schwartzman,1999 2002). In this modeling abiotic $W_{sil}$ corresponds to $G = 0$.

At 15 °C $BEW_o = BG_o^{0.4} + 1 = 1.23B + 1$. For example, at 15 °C assuming $B = 20$ gives $BEW = 25.6$.

Global biological productivity (G) is given by

$$(6) \quad G = 2\,\exp[-\sum k^{-1}\,((T-25)/25)^{2k}\,] \quad \text{with the sum over k from 1 to 5.}$$

This parameterization of biologic productivity as a function of temperature is modified from the parabolic representation, i.e., $G = 2(1 – ((T – 25)/25)^2)$, where $G = 2$ at 25 deg C, and $G = 0$ at 0 and 50 °C (Caldeira and Kasting, 1992; Von Bloh et al., 2003). Our formulation is a bell shaped curve that follows the parabola closely for temperatures between 0 and 50 °C but approaches 0 asymptotically as temperatures go beyond these bounds. This avoids the problem of negative productivity values at high temperatures that cause singularities in the differential equations. We have assumed here that biotic productivity in the ocean tracks the variation of biotic productivity on land. This assumption is supported by the dependence of marine productivity to the weathering
release of phosphorus from land (Tyrell, 1999; Filipelli, 2010; Planavsky et al., 2010).

For organic carbon burial ($F_3$) we have

(7)    $F_3 = F_{3o} \times (G/G_o)^2 = 0.3543\, F_{3o} \times G^2$

$G_o$ is the value of G at 15 °C, $G_o = 1.680$ (37).

Oxidation of reduced organic carbon ($F_4$) is

(8)    $F_4 = F_{4o} \times e^{b(T-15)}$,

where b ranges from 0.056 to 0.0705. The low value of b corresponds to the abiotic factor as in the case of silicate weathering and the high value corresponds to the factor estimated for heterotrophic respiration (Anderson-Teixeira et al., 2008). We assume that given the presence of free oxygen in the atmosphere, the oxidation rate of reduced organic carbon does not depend on the oxygen level (Berner, 1999; Bolton et al., 2006).

**Greenhouse functions**

The first greenhouse function we used was

(9)    $T = 4.55\,(Pr^{0.364} - 1) + 15$    or solving for Pr:    $Pr = [((T - 15)/4.55) + 1]^{2.747}$.

As previously defined, T is °C,

$Pr = (pCO_2 / pCO_{2\ reference})$, $pCO_2$ is the atmospheric carbon dioxide level in bars, reference state is 0.0003 bars.

This is a simplified greenhouse function (Schwartzman, 1999 2002; Walker et al., 1981). It assumes the present solar luminosity, so an adjusted greenhouse function is needed for simulations of earlier geologic time such as the Paleozoic. Further, it assumes the planetary albedo is constant.

A second greenhouse function (Caldeira and Kasting, 1992a) was used as well in a sensitivity analysis. In this formulation the difference between the mean global surface temperature and the effective radiating temperature of the earth is modeled. The mean global surface temperature *in Kelvin* is $T$, and $T_{rad}$ is the effective radiating temperature of the earth (in our model we assume this to be a constant 255 K). The difference is $\Delta T$ so that

(10)   $T = T_{rad} + \Delta T$, where $T$ is the mean global surface temperature in Kelvin (K).

$\Delta T = 815.17 + (4.895 \times 10^7)T^{-2} - (3.9787 \times 10^5)T^{-1} - 6.7084y^{-2} + 73.221y^{-1} - 30{,}882\,T^{-1}y^{-1}$,

where $y = \log_{10} pCO_2$ and $pCO_2$ is in bars). Equation (10) is valid for $pCO_2 < 0.03$ equivalent to $Pr < 100$. It apparently gives more plausible temperatures for the low pCO2 range than equation (9)), at least for Pr less than roughly 30 (Figure 2).   The two greenhouse functions are shown in Figure 2.

Note that the extrapolated greenhouse function consistent with IPCC estimates, with $pCO_2$ sensitivity of 4 (PALAEOSENS Project Members, 2012), falls between the curves corresponding to equations (9) and (10) except very close to the upper limit for plants, 50 ºC. There is proposed evidence for high climate sensitivity during the Cretaceous and Cenozoic based on paleotemperatures/pCO$_2$ levels (Royer et al., 2012).

The greenhouse functions used assume the planetary albedo remains constant, equal to the present value. It is plausible that the planetary albedo itself may be variable in the postulated climate states in the temperature range assumed here, potentially another variable to be included in further modeling.

**Additional explanation of the sensitivity analysis**

Keeping V = 0.5 while T is changing, with R a function of T, the departure of the system from steady-state (V = R) is given by

(11) $dpCO_2/dt = V - R = 0.5 - 0.5\,(BG^{0.4} + e^{0.073(T-15)}\,Pr^{\alpha})/(1.23B + 1) - 0.0354G^2 + 0.1\,e^{b(T-15)}$

The parameter B magnifies the weathering flux as a result of a given level of biological productivity, and is essential in determining the existence of higher temperature steady states. It's therefore necessary to explore the relationship between B and the higher temperature steady states. The steady state condition obtains when V = R and equation (11) equals zero.

(12)   $0 = 0.5 - (0.5\,(BG^{0.4} + e^{0.073(T-15)}\,Pr^{\alpha}))/(1.23B + 1) - 0.0354G^2 + 0.1\,e^{b(T-15)}$

Solving for B gives

(13)   $B = (2Q - e^{0.073(T-15)}\,Pr^{\alpha})\,/\,(G^{0.4} - 2.46Q)$ where $Q = 0.5 + 0.1\,e^{b(T-15)} - 0.0354G^2$

This equation is graphed for values of α and b in Figure 3c using equation (9), and in Figure 3d using equation (10). Equation (13) is the "bifurcation curve", i.e. the curve that describes the possible steady-states of T for varying values of B. This equation cannot be explicitly solved for the steady-state values of T but can be plotted by switching the axes on the graph T as dependent on B. For fixed values of $F_{3o}$ and $F_{4o}$ (e.g. $F_{3o} = F_{4o} = 0.1$), B is a "U" shaped function of T. When the x and y axes are flipped the curve rests on its side, opening to the right. The upper branch corresponds to the stable and the lower branch the unstable steady-states respectively (Figures 3c and 3d). The critical point of the curve corresponds to the left most point (and the lowest temperature where a stable steady-state exists), and is derived by setting the derivative of equation (13) to zero (this was done numerically). Figure 3e was obtained by allowing $F_{3o}$ and $F_{4o}$ to vary, along with $W_{sil}$, so that all values lie in the interval [0, 1], again assuming V equals 0.5, producing a stable steady-state at 15 ºC. The critical values of T were thereby mapped over these combinations of $F_{3o}$ and $F_{4o}$ to produce the UTSS surface shown in Figure 3e. Results were computed using equation (9).

**Modeling early Earth climates, considering the potential variation of planetary albedo**

As a first approximation, before the emergence of oxygenic photosynthesis, we assume significant organic carbon formation and oxidation can be neglected. We also note that prior to the rise of atmospheric oxygen at the end of the Archean the weathering of ferrous iron silicates contributed to the removal of carbon dioxide from the atmosphere/ocean because of the solubility of $Fe^{+2}$ and its precipitation on the ocean floor as Fe carbonate, but this flux can be neglected in the Phanerozoic.

In previous modeling of long-term carbon cycle the planetary albedo has been assumed to be constant in greenhouse functions, or a function of surface temperature, because of Rayleigh scattering, especially at high $pCO_2$ levels (Kasting and Ackerman, 1986; Caldeira and Kasting, 1992b). Our modeling of the Phanerozoic assumes constant planetary albedo, similar to the present. This assumption is supported by the fact that empirically derived paleotemperatures and paleo $pCO_2$ levels are consistent with model predictions using a greenhouse function with a constrained low planetary albedo similar to the present (Berner and Kothavala, 2001). But is this also true of potentially more extreme climate regimes such as in the Archean?

A potential major influence on planetary albedo is the potential variation of cloud cover/behavior as a function of $pCO_2$, and temperature. Will this dependence lead to multiple potential steady-states? For example, high T, low $pCO_2$ levels and low planetary albedo and conversely low T, high $pCO_2$ levels and high planetary albedo could generate the same weathering forcings, equal to the same volcanic source at stable steady-states. Figure 4 shows schematic hypothetical potential steady-states corresponding to the intersections of linear greenhouse curves with a constant weathering forcing, labeled for example as a, b, c, d. The weathering forcing corresponds to $W_{sil}$ in equation (5) minus the biotic component. Each greenhouse curve corresponds to a different planetary albedo. These potential steady-states may have implications to paleoclimates of the Hadean and early Archean, prior to the rise of oxygenic photosynthesis, when biotic productivity was likely limited compared to more recent times. We look forward to studies of the long-term carbon cycle using the 3D GCM approach on paleoclimates, an exciting advance recently applied to the early Earth (Charnay et al., 2013).

**Figure Legends**

**Figure 1. Long-term Carbon Cycle.** All fluxes correspond to carbon dioxide transfer. V is the volcanic outgassing flux, $F_1$ corresponds to deposition flux of CaMg carbonates in ocean, $F_2$ to the weathering flux of carbonates on land, $F_3$ to the burial flux of reduced organic carbon $F_4$ to the oxidation flux of reduced organic carbon in sediments on land.

**Figure 2. Greenhouse functions, Temperature (ºC) versus Pr:** Temperature corresponds to the global mean surface value. Pr = ($pCO_2$ / $pCO_2$ pre-anthropogenic modern level, taken as 0.0003 bars), $pCO_2$ being the atmospheric carbon dioxide level in bars. Equations (9) "Walker" and (10) "C & K" are so labeled. Extrapolated greenhouse function with $CO_2$ sensitivity of 4 from ref. 41.

**Figure 3. Sensitivity Analysis of Model Results**

**a. (V - R) versus Temperature (ºC).** (V – R) is the measure of departure from steady-state. This schematic graph illustrates stable (labeled S) and unstable (labeled U) steady-states.

**b. (V - R) versus Temperature (ºC).** (V – R) is the measure of departure from steady-state. Curves generated with B varying between 27.409 and 30.293, $\alpha = 0.3$ and b = 0.056. S is a stable steady-state, U an unstable steady-state. Greenhouse function: equation (9); see Supplementary Information for definitions of B, $\alpha$ and b.

**c. Temperature (ºC) versus B.** Stable and unstable steady-states shown for limiting values of $\alpha$ and b. Greenhouse function: equation (9); see Supplementary Information for definitions of B, $\alpha$ and b.

**d. Temperature (ºC) versus B.** Stable and unstable steady states shown for limiting values of $\alpha$ and b. Greenhouse function: equation (10); see Supplementary Information for definitions of B, $\alpha$ and b.

**e. Upper Temperature Steady-States** shown as values on contoured temperature surface for varying values of $F_{3o}$ and $F_{4o}$, assuming again a stable steady-state at 15 ºC with V held constant at 0.5. $F_o$ is the flux assumed at 15 ºC. Greenhouse function: equation (9); see Supplementary Information for definitions of $\alpha$ and b.

**f. $\alpha$ versus b contour plot** showing 2nd stable steady-state. Contours give the lowest temperature that the 2nd stable steady-state can exist for a particular combinations of $\alpha$ and b. Greenhouse function: equation (9).

**g. Values of log ($P_r$) versus Temperature (ºC) and B.** Steady-states are sketched on the surface as lines, using Greenhouse functions "Walker" (equation 9) as red and "C&K" (equation 10) as black, with $\alpha = 0.30$ and b = 0.056.

**Figure 4. Log $pCO_2$ versus Surface Temperature for early Earth climates.** Source Flux V and Weathering Forcing (WF) assumed constant and equal, with different greenhouse curves corresponding to variation in the planetary albedo. Surface temperature and $pCO_2$ levels in same units as in Figure 2.

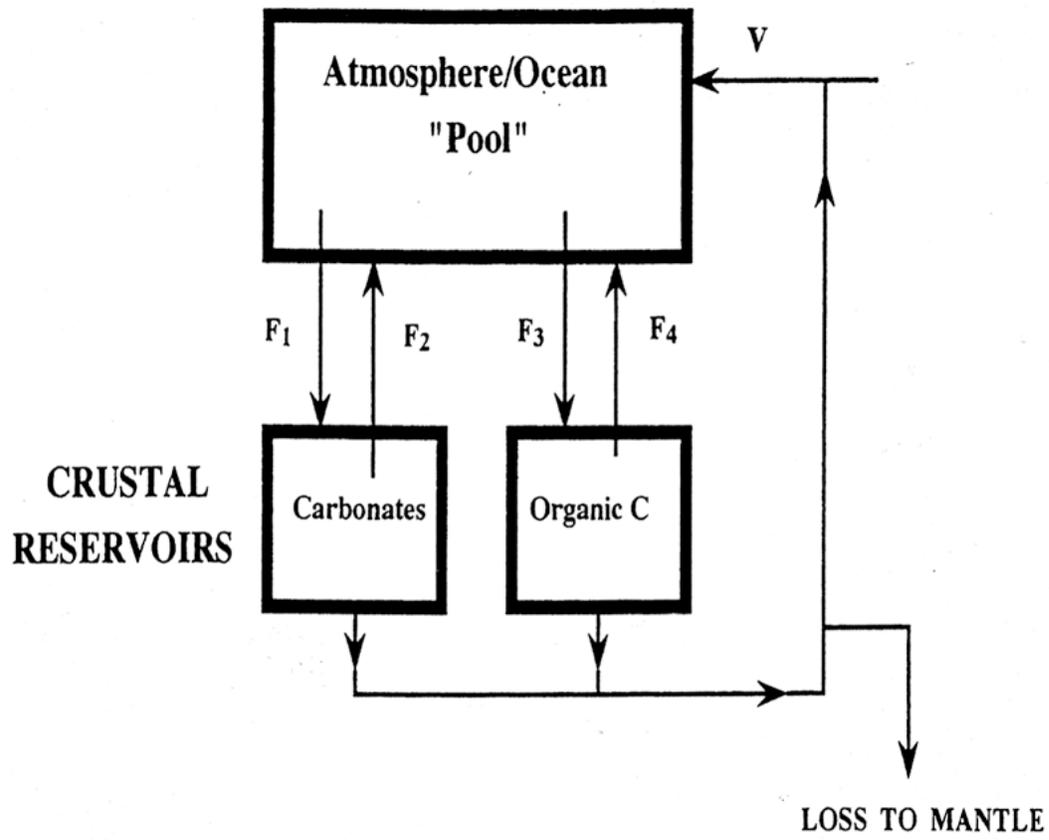

**Figure 1**

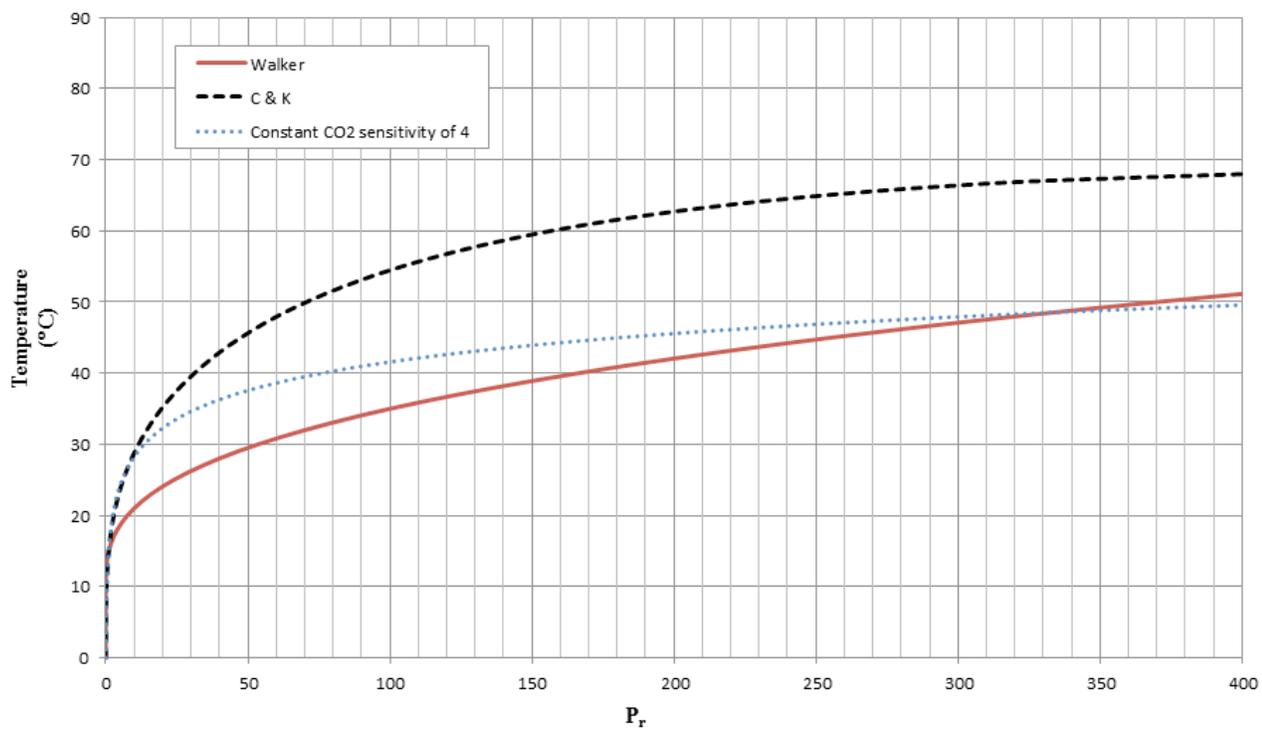

**Figure 2**

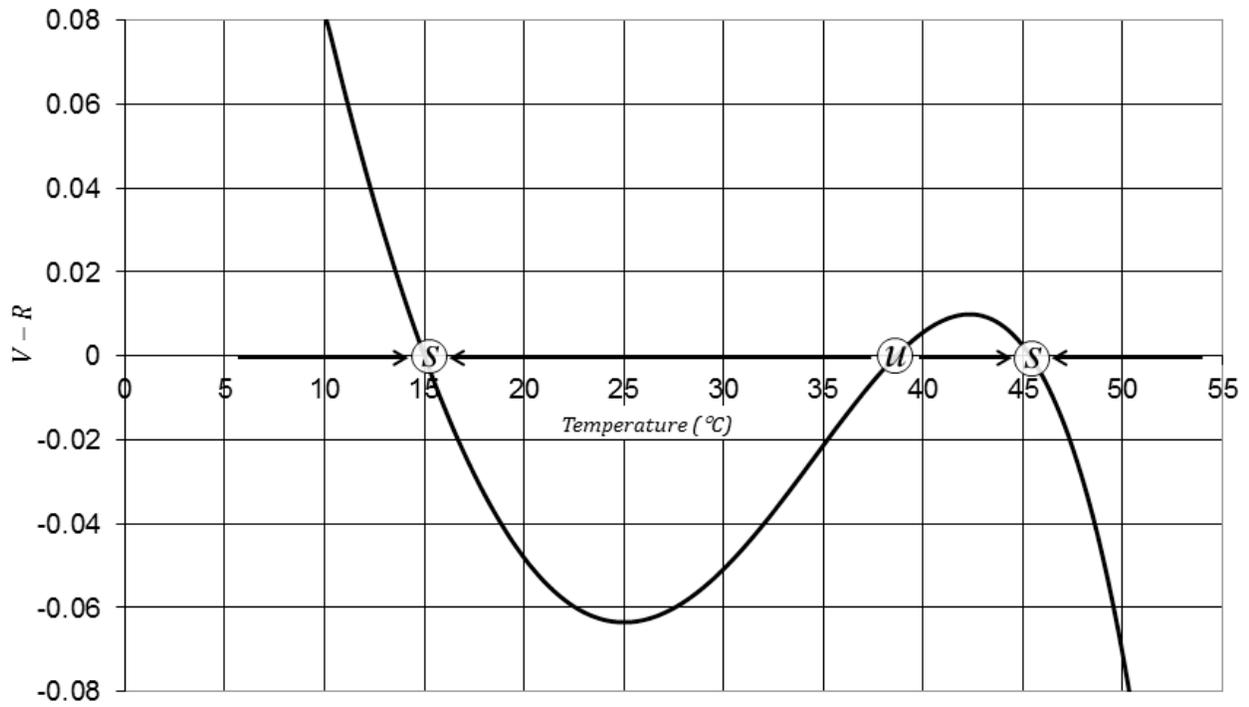

**Figure 3a**

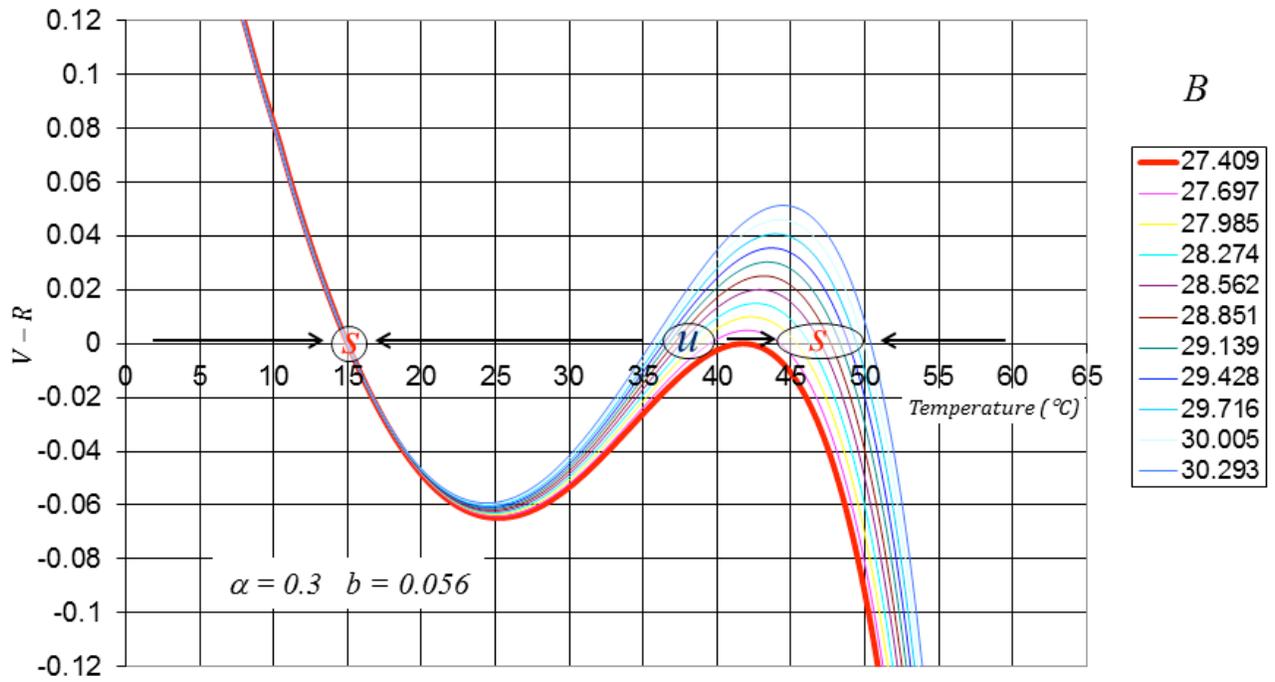

**Figure 3b**

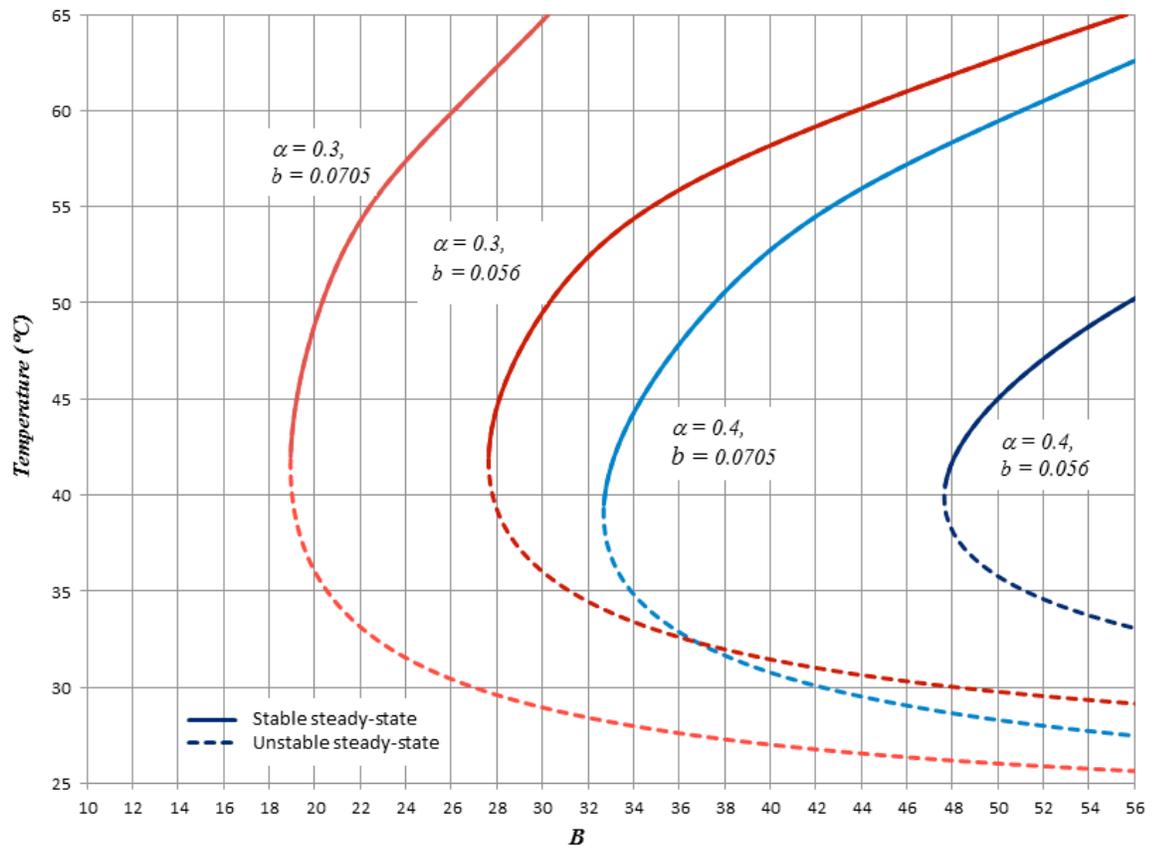

**Figure 3c**

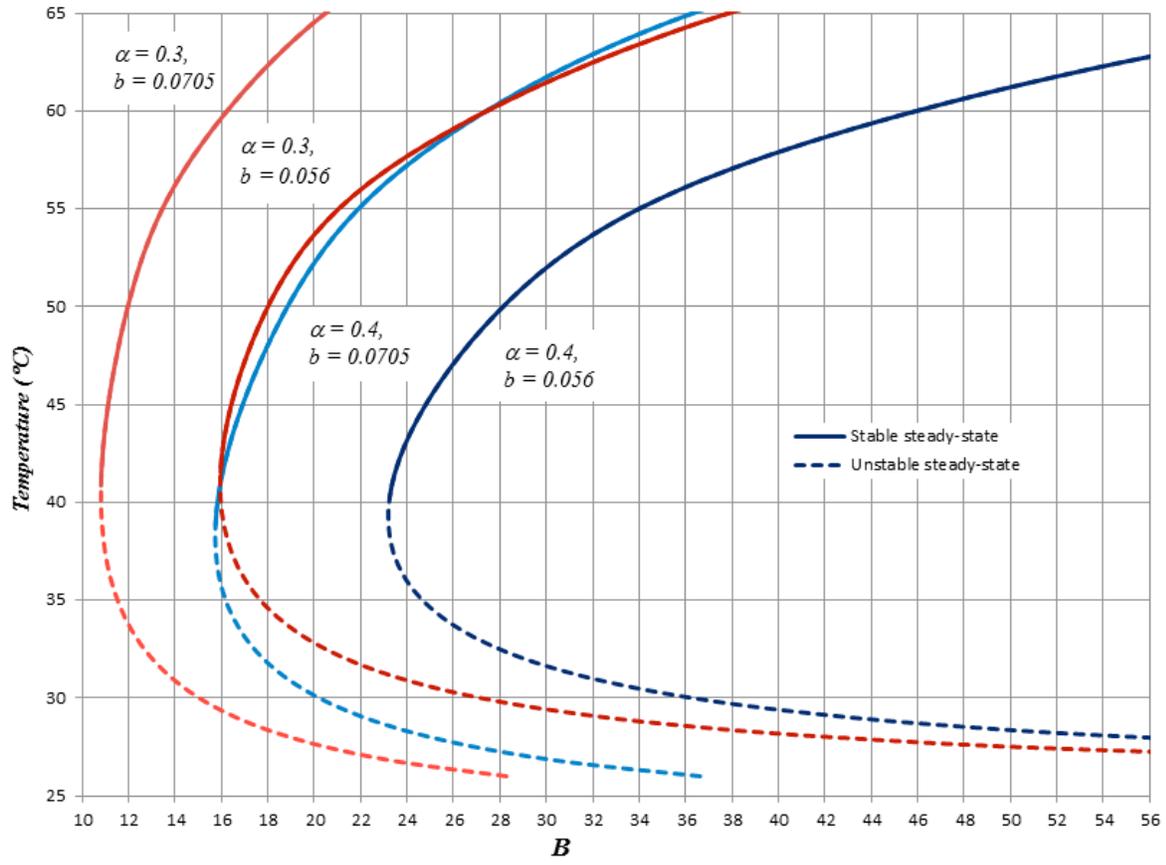

**Figure 3d**

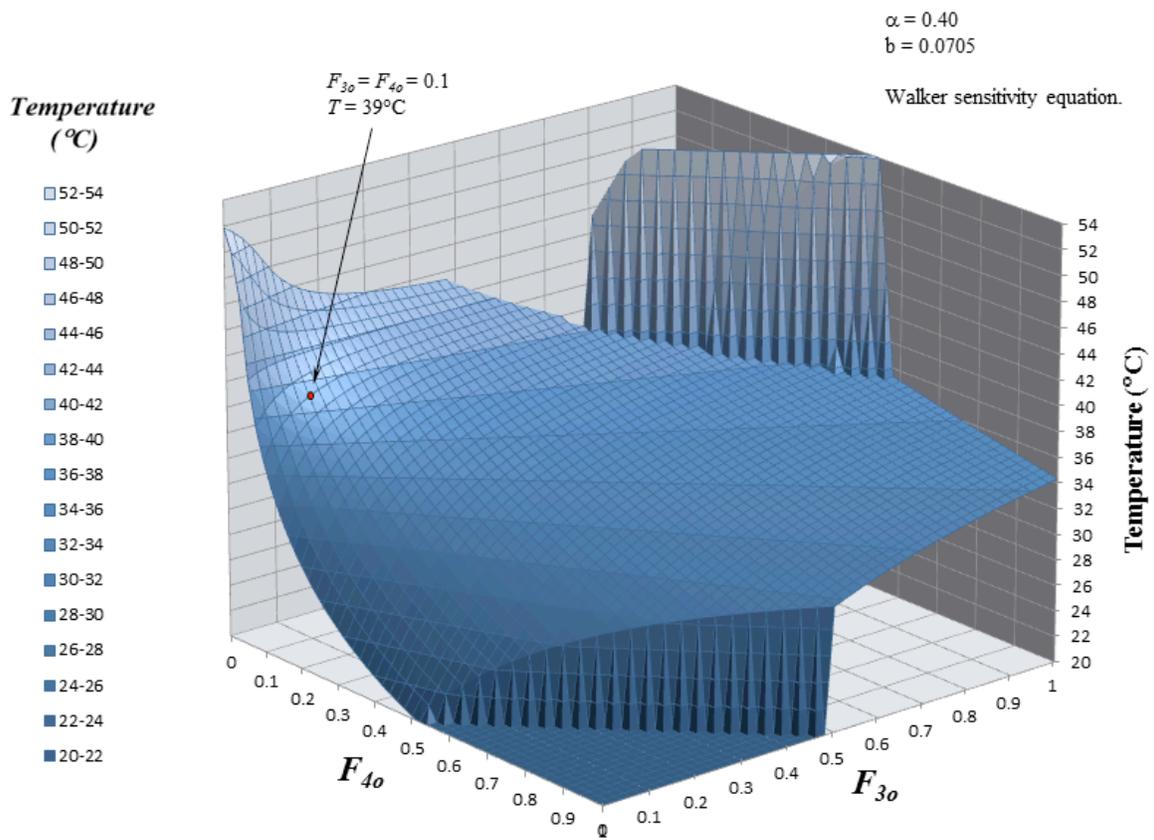

**Figure 3e**

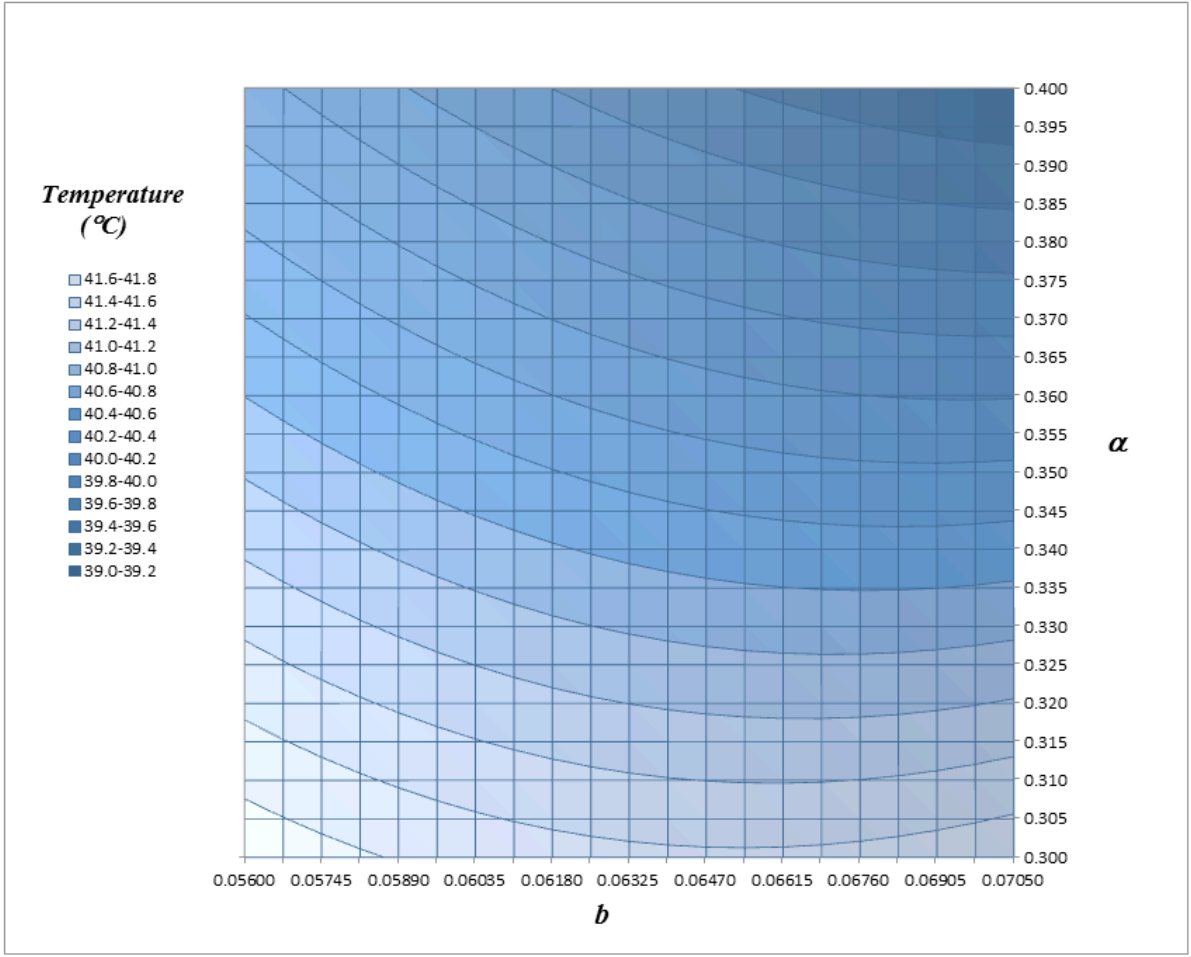

**Figure 3f**

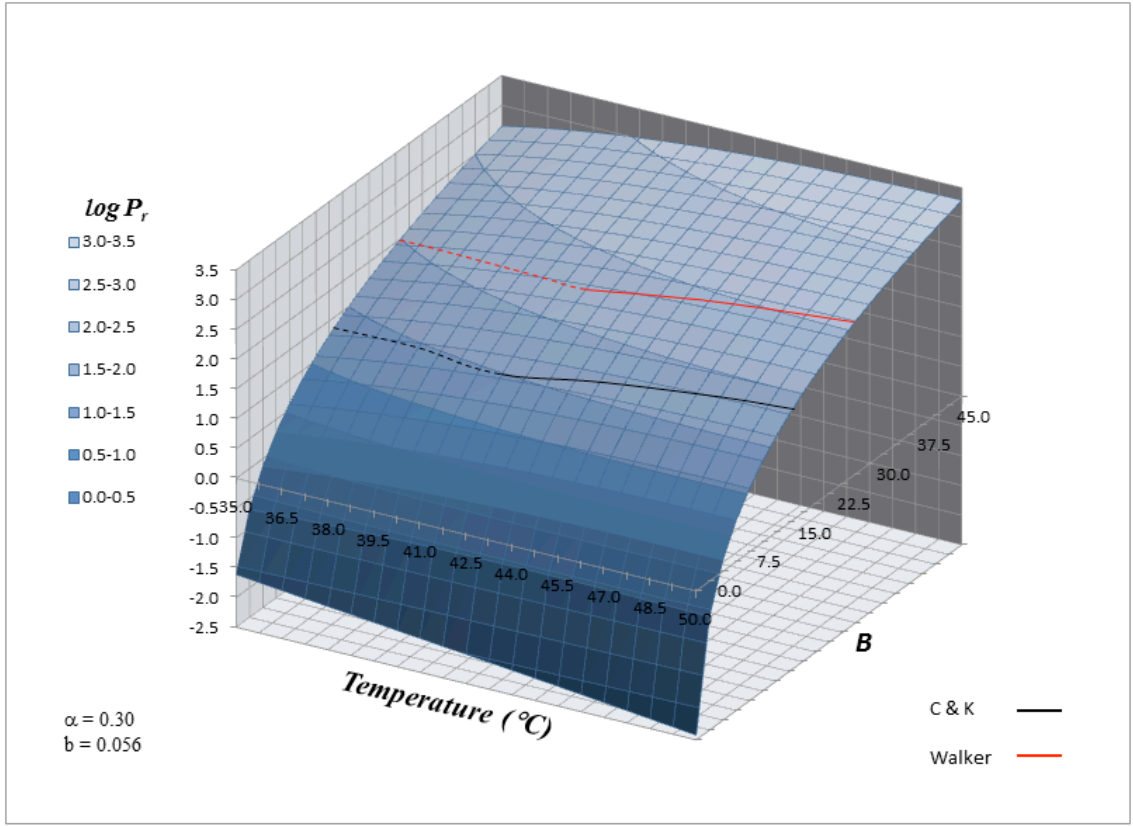

**Figure 3g**

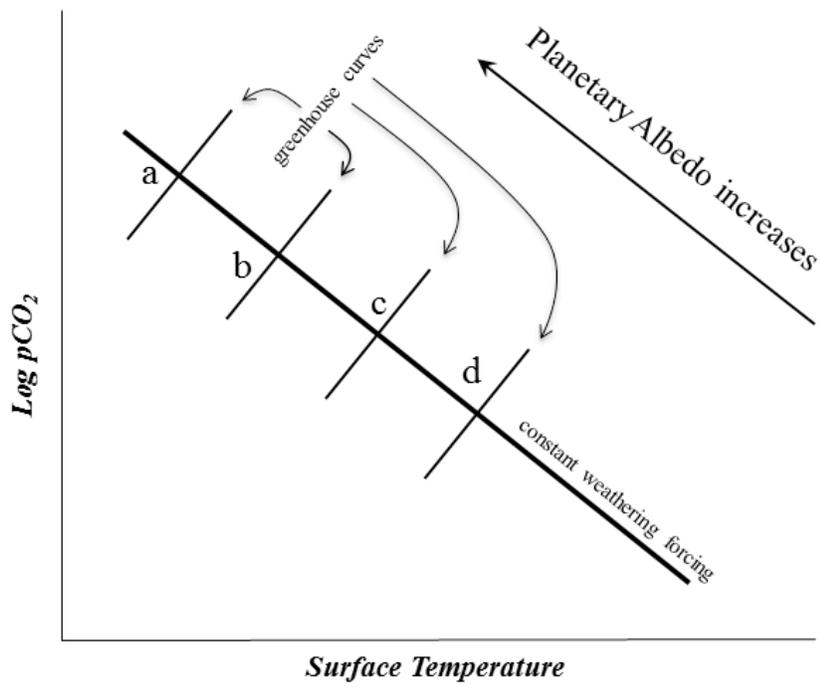

**Figure 4**